\documentclass[journal=jpcbfk,manuscript=article,groupaddress,layout=twocolumn]{achemso}
\usepackage{natbib}

\renewcommand{\vec}[1]{\mbox{\boldmath$#1$}}

\newcommand{\dif}{\mathrm{d}}

\newcommand*{\citen}{}
\DeclareRobustCommand*{\citen}[1]{%
  \begingroup
    \romannumeral-`\x 
    \setcitestyle{numbers}%
    \cite{#1}%
  \endgroup
}

\usepackage{graphicx}

\usepackage{amsmath}

\usepackage{amsfonts}

\usepackage{graphicx} 

\let\oldmaketitle\maketitle
\let\maketitle\relax

\usepackage{float}

\title{Statistical analyses of hydrophobic interactions:  A mini-review}
\author{L. R. Pratt}
\email{lpratt@tulane.edu}
\affiliation{Department of Chemical and Biomolecular Engineering, Tulane
University, New Orleans, LA 70118}

\author{Mangesh I. Chaudhari}
\email{michaud@sandia.gov}

\author{Susan B. Rempe}
\email{slrempe@sandia.gov}
\affiliation{Center for Biological and Engineering Sciences, Sandia National Laboratories,  
Albuquerque, NM 87185}

\begin{document}

\date{\today}

\twocolumn[ \begin{@twocolumnfalse} \oldmaketitle \begin{abstract} This review
focuses on the striking recent progress in solving for hydrophobic
interactions between small inert molecules. We discuss several new
understandings. Firstly, the \emph{inverse temperature} phenomenology of
hydrophobic interactions, \emph{i.e.,} strengthening of hydrophobic bonds with
increasing temperature, is decisively exhibited by hydrophobic interactions
between atomic-scale hard sphere solutes in water. Secondly, inclusion of
attractive interactions associated with atomic-size hydrophobic reference cases
leads to substantial, non-trivial corrections to reference results for purely
repulsive solutes. Hydrophobic bonds are \emph{weakened} by adding solute
dispersion forces to treatment of reference cases. The classic statistical
mechanical theory for those corrections is not accurate in this application, but
molecular quasi-chemical theory shows promise. Finally, because of the
masking roles of excluded volume and attractive interactions, comparisons that
do not discriminate the different possibilities face an interpretive danger.
\end{abstract} \end{@twocolumnfalse}]

\maketitle

\section{Introduction}

The molecular theory of hydrophobic effects,\cite{Chan:79} and
particularly of hydrophobic interactions,\cite{franks1975hydrophobic} has been a
distinct intellectual challenge for many decades. The intellectual challenge
originates with the entropy-driven character of hydrophobic
interactions.\cite{Tucker:1979wk,RosskyPJ:BENIIA,Pratt85} Explanation of
those entropies requires molecular statistical mechanics. Statistical
mechanical theories of hydrophobic effects have been rare,  difficult, and
unconvincing in their inceptions,\cite{Chan:79,PrattLR:Molthe}
especially in contrast to the voluminous and graphic results from molecular
simulations of aqueous solutions of hydrophobic species.  

It is correct and traditional to introduce work on hydrophobic effects by
calling-out their vast importance.\cite{berne2009dewetting} But the subject is
so vast that the traditional calling-out scarcely ever does justice to the
whole. We submit to that situation and pattern here, giving a few examples, and
citations of discussions more extended than would fit in this mini-review. 
Hydrophobic effects can play a prominent role in practical materials science associated with aqueous
solution interfaces.
\cite{kumar1994patterning,de1999hydrophobic,
giessler2004hydrophobic,guo2005stable,cygan,Brinker:2008} 
Work on clathrate hydrates, in which molecular cages of water
trap gases,\cite{clawson:h2,Koh2016} provides an example pertinent to energy applications.
More traditional yet is work on biomolecular
structure,\cite{richards1977areas,ISI:A1991EP16300008} and even more expansive
discussions are associated with the origins of
life.\cite{TANFORDC:Thehea,Pohorille:2012ki} We note that an immediate precedent
of this work offered a focused discussion of the expression of hydrophobic
interactions in aqueous polymer solutions,\cite{Chaudhari:2016gm} and closely
related work gave an extended discussion of theoretical work on atomic-scale
hydrophobic interactions.\cite{Asthagiri:2008in}
 
The subjects of hydrophobic hydration and hydrophobic interactions are indeed
vast, and this review is not exhaustive. Instead we focus on the recent
interplay between statistical mechanical theory of hydrophobic interactions and
simulation experiments designed to test and clarify those theories.

\begin{figure}
  \begin{center}
\includegraphics[width=3.0in]{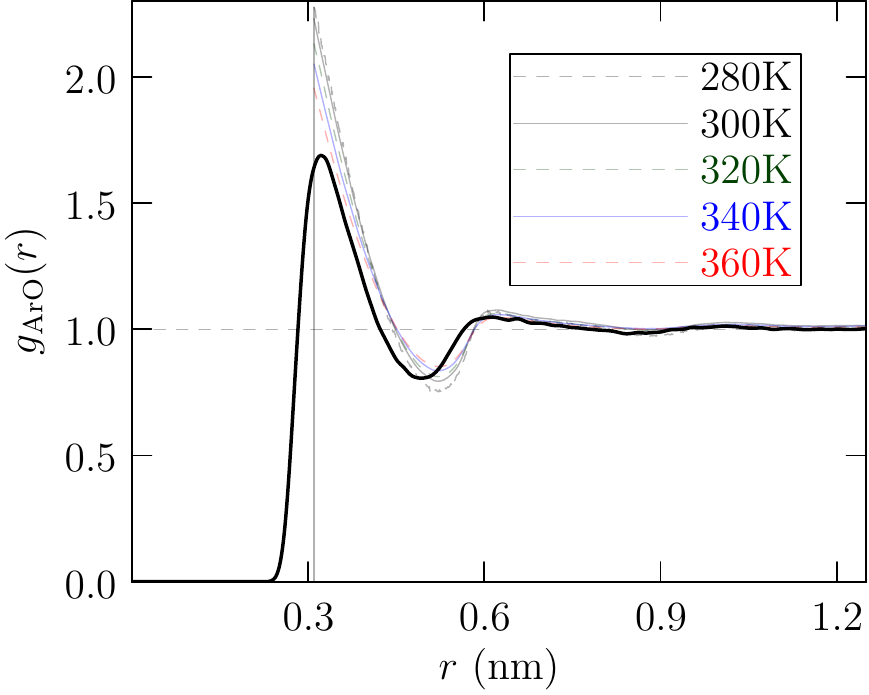}
\caption{Radial distribution of water O atoms from a dissolved Ar
atom, $T$ = 300~K, $p$ = 1~atm (heavy curve). Radial distributions (fainter,
background curves) for hard-sphere model solutes with distances of closest
approach, $\lambda = 0.31$~nm, on the basis of cavity methods
\cite{chaudhari2013molecular,Palma92,MICThesis}.}
\label{fg:gAO} 
  \end{center}
\end{figure}

The distinction of hydrophobic interactions from hydrophobic hydration is
important and well-recognized, but deserves emphasis. 
Hydrophobic interactions are free energy changes resulting 
from the mutual positioning of hydrophobic species in water. 
In contrast, hydrophobic hydration addresses the placement of the 
water molecules neighboring hydrophobic species in water. An example of a
hydrophobic hydration characteristic is the radial distribution, $g_{\mathrm{ArO}}(r)$, of water oxygen (O)
atoms near an argon (Ar) atom solute in water (FIG.~\ref{fg:gAO}). This shows
crowding of water near that hydrophobic solute. Water structures
similarly around other gases, including hydrogen
\cite{Kirchner:2004,dsabo:06,dsabo:08,Smiechowski:2015}
and even carbon dioxide.\cite{jiao} 
Hydrophobic interactions and
hydrophobic hydration phenomena are expected to be fundamentally related. But
they have shown distinctly different variabilities, and theories of hydrophobic
hydration\cite{Rogers2012} 
have gone further than theories of hydrophobic interactions.   

\begin{figure}
\includegraphics[width=3.25in]{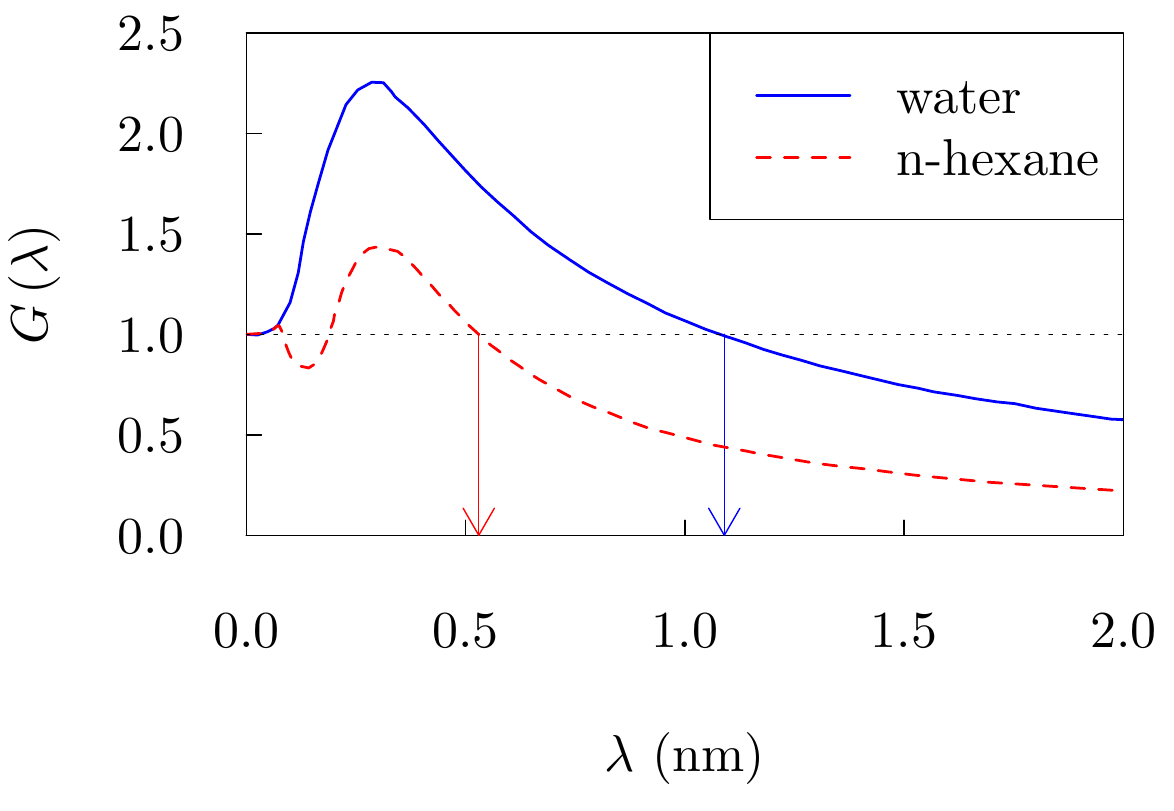}
\caption{For a hard sphere in water and $n$-hexane, contact values of
radial distribution of O (water) or C ($n$-hexane) from application of the
revised scaled-particle model. Results for $T= 270$~K along the saturation
curves of water and hexane, redrawn from Ref.~\citen{hashb07}. The maximum value for
water, about 2.2, agrees with the radial distributions of FIG.~\ref{fg:gAO}
though the precise values of $\lambda$ near 0.3~nm are slightly different. Both
water and $n$-hexane cases show  density depletion in  the local neighborhood of
the hard sphere for gradually larger $\lambda$, but this behavior is more advanced for
$n$-hexane: $G(\lambda) < 1$, for $\lambda > 0.53$~nm for $n$-hexane,
and for $\lambda > 1.09$~nm for water. }
\label{fg:one}
\end{figure}

Another hydrophobic hydration characteristic, and one for which there is well-developed
statistical mechanical theory, is the dependence of the contact value of the
hard-sphere radial distribution functions (FIG.~\ref{fg:gAO}) on the distance,
$\lambda$, of closest approach of a water O atom (FIG.~\ref{fg:one}). The
maximum of $G(\lambda )$ in FIG.~\ref{fg:one} provides an unambiguous
separation of large and small $\lambda$ regimes. Here that boundary is near
$\lambda = 0.3$~nm. $G(\lambda )$ supplies the compressive force exerted by the
water on a hard cavity.\cite{PrattLR:Thehea}  By falling in the neighborhood of
that maximum, the results of FIG.~\ref{fg:gAO} demonstrate the case of the \emph{strongest
compressive force} on a hydrophobic spherical exclusion. The results
(FIG.~\ref{fg:one}) also draw a clear distinction between solvation structuring
of $n$-hexane, a typical organic solvent, and liquid water.

\begin{figure}
  \begin{center}
\includegraphics[width=3.0in]{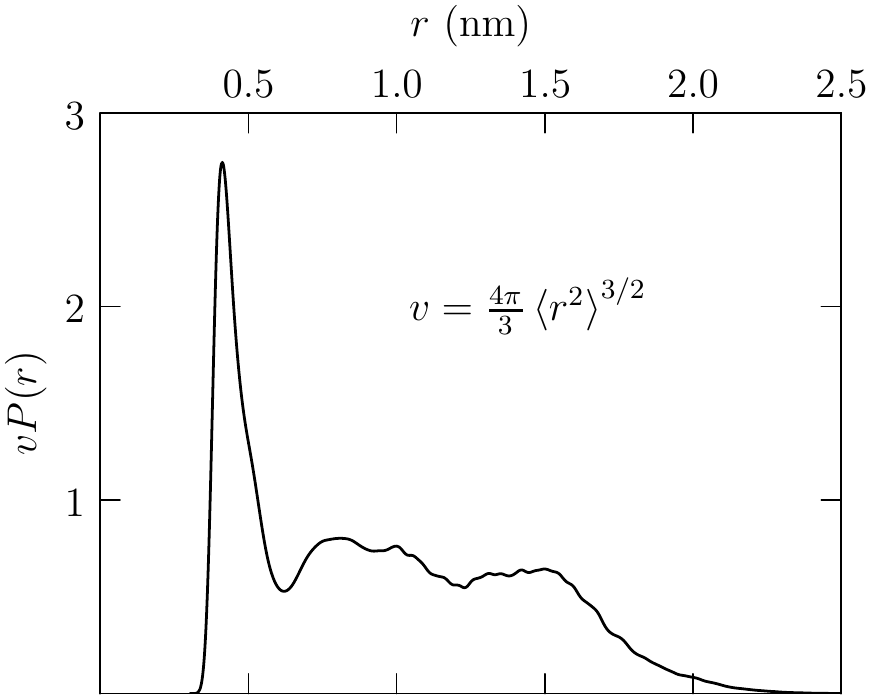}
\includegraphics[width=3.0in]{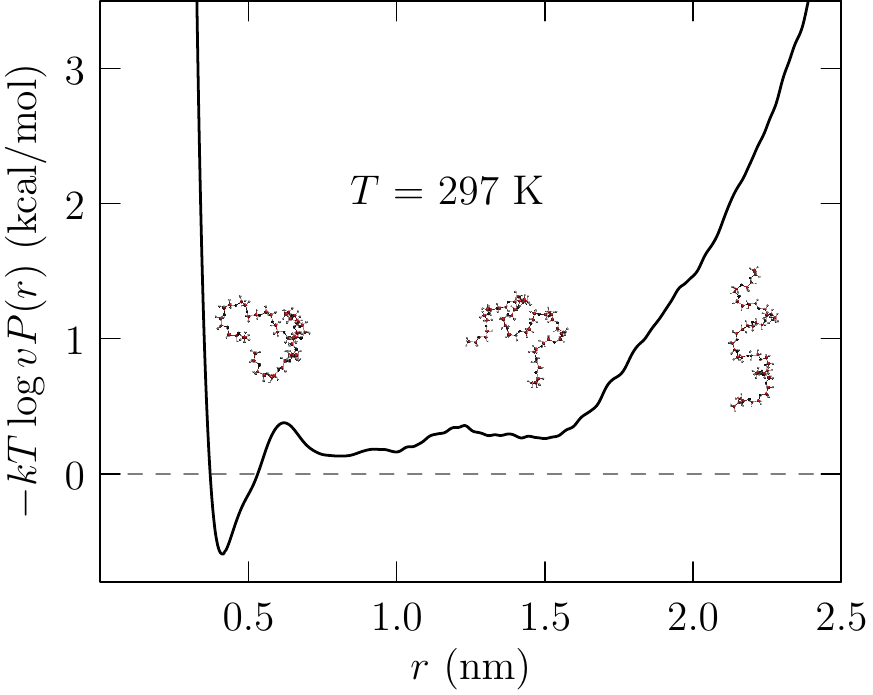}
\end{center}
\caption{(upper) Probability density $P(r)$ for methyl-methyl length for
[CH$_3$(CH$_2$-O-CH$_2$)$_m$CH$_3$](aq) with $m$ = 21. The normalization is
chosen in analogy with conventional atom-atom radial distribution functions of
liquids. (lower) Potential of the average end-to-end forces showing distinct
loop-closure, globule, and high-extension regions. The choice of normalization
for the upper panel also sets the origin of the $y$-axis of the lower panel.
Results for $r < 1.0$~nm were obtained with the WHAM procedure.\cite{Kumar1992} Those 
high-resolution results were matched to observation of $P(r)$ overall from parallel
tempering molecular simulations \cite{chaudhari_communication:_2010}.
\label{fig:fig2}}
\end{figure}

In contrast with hydrophobic hydration, the association (FIG.~\ref{fig:fig2}) of
end-methyl groups capping a poly-ethylene-oxide (PEO) oligomer shows a distinct
hydrophobic interaction \cite{chaudhari_communication:_2010,Chaudhari2014}. 
That interaction is also suggested by the density profile of 
poly-dimethylsiloxane (PDMS) interacting with a solid surface.\cite{Tsige:2003}
The PEO example was constructed for feasible direct measurement, \emph{e.g.,} by neutron
diffraction, of the hydrophobic interactions shown. Preliminary experiments of
that sort, studying end-labeled PEO chains, are already available
\cite{MICThesis}. Simpler experimental studies of loop-closure
\cite{Weikl:2008ii} of PEO chains were later discussed as an extension of this
idea.\cite{chaudhari2014loop}

This review focuses on the striking recent progress in solving for these
hydrophobic interactions between small inert molecules
\cite{Koga:2013el,chaudhari2013molecular,Chaudhari:2015jq,Ashbaugh:2015cxa,Chaudhari:2016gm}.
We anticipate discussion below by noting the new understandings. (1) The
\emph{inverse temperature} phenomenology of hydrophobic interactions,
\emph{i.e.,} strengthening of hydrophobic bonds with increasing temperature, is
decisively exhibited by hydrophobic interactions between atomic-scale hard
sphere solutes in water.  (2) Inclusion of attractive interactions associated
with atomic-size hydrophobic reference cases leads to substantial corrections to
reference results for purely repulsive solutes. Hydrophobic bonds are
\emph{weakened} by adding solute dispersion forces to treatment of reference
cases \cite{Asthagiri:2008p1418,Chaudhari:2016gm}. The classic statistical
mechanical theory for those corrections is not accurate in this application
\cite{Asthagiri:2008p1418,Chaudhari:2016gm}, but molecular quasi-chemical theory (QCT)
shows promise \cite{Asthagiri:2008p1418}. (3) Theories, comparisons, models, or
pictures that do not distinguish effects of reference (repulsive force) cases
from those with interactions, including masking attractions, are not safe
\cite{PANGALIC:AMCs,ghumm96,Garde:1996p7972}.   

In closing this Introduction, we
point\cite{PrattLR:Hydeam,Pohorille:2012ki,pratt2007special} to the most
ambitious attempt to extract physical insight from the statistical mechanical
theories that work for hydrophobic
effects.\cite{Pierotti:1976tg,Stillinger:73,Gomez:1999hm,AshbaughHS:Colspt} That
analysis suggested that the equation of state of liquid water is the primary
source of peculiarity of hydrophobic
effects.\cite{Pohorille:2012ki,pratt2007special,AshbaughHS:Colspt} The
compressibility of water is low, compared to organic solvents. Water is stiffer
and that low compressibility is weakly sensitive to temperature along the vapor
saturation curve. Similarly, the density of water changes unusually slowly along
that vapor saturation curve. Good theories should faithfully incorporate those
equation of state characteristics even if they compromise the molecular-scale
description of hydration structure. This ``equation of state'' explanation of
hydrophobic effects may not be fully sufficient on its own, but it is the most
defensible picture currently available.

\section{Cavity Methods to Obtain Hydrophobic Interactions between Hard-spheres 
in Water} 

\begin{figure}
\includegraphics[width=3.0in]{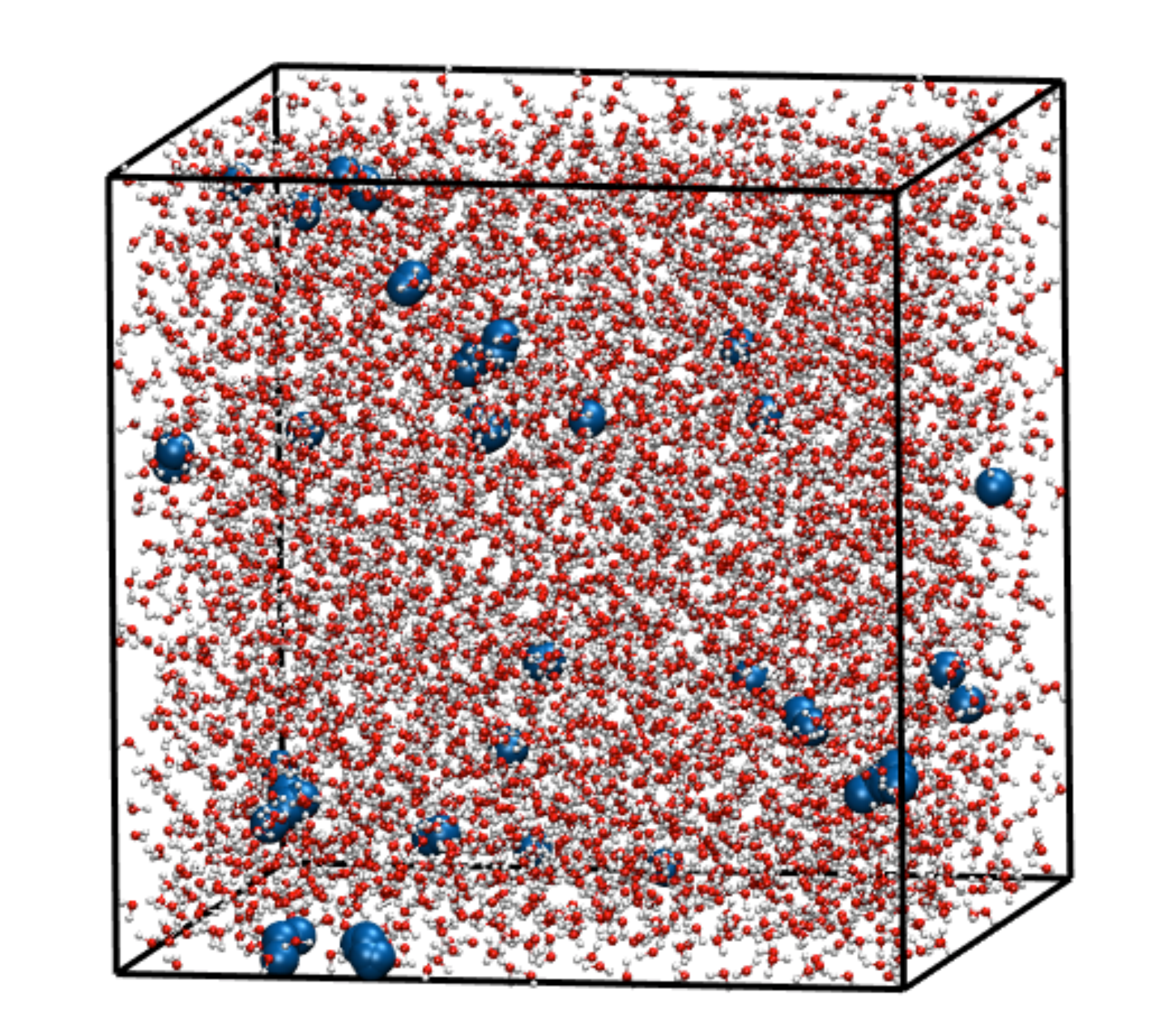}
\caption{A configuration of 5$\times 10^3$ water molecules together with the
spherical inclusions identified by $n_\mathrm{t}$ = 2$\times 10^5$ trial
placements of a hard sphere with distance of closest approach to an oxygen atom
of 0.31~nm. This size corresponds approximately to A=Ar solute, for which the van
der Waals length parameter $\sigma_\mathrm{A}$ is about 0.34~nm \cite{lowHill}.
Thus, we adopt 0.31~nm - 0.17~nm = 0.14~nm as a van der Waals contact radius of
the water oxygen atom. Hard sphere solutes of this size have about maximal O
contact density (FIG.~\ref{fg:one})\cite{hashb07,AshbaughHS:Colspt}. }
\label{fig:eight} 
\end{figure}

We first discuss hydrophobic interactions between hard-sphere solutes, A
\cite{chaudhari2013molecular}. We consider the radial distribution function  
\begin{eqnarray} g_{\mathrm{AA}}(r) = \exp{\left\lbrack
- \beta u_{\mathrm{AA}}(r)\right\rbrack }y_{\mathrm{AA}}(r)
\label{eq:y}
\end{eqnarray} 
for atomic-size hard spheres relying on the \emph{cavity distribution function,}
$y_{\mathrm{AA}}(r)$, exploiting trial placements into the system volume
(Fig.~\ref{fig:eight}). We make $n_t$ trial placements into the system volume
$V$ for each configuration sampled by the molecular simulation. Those placements
are spatially uniform, and $\left(n_\mathrm{t}-1\right)\Delta V
/V$ should land in a volume $\Delta V$, which is a thin shell of
radius $r>0$ surrounding a permissible insertion.  The number of permissible 
placements obtained in the shell for
each configuration is denoted by $\Delta
n_\mathrm{s}(r)$.   Averaging over configurations, we estimate
\begin{eqnarray} \left(\frac{ \overline{n}_\mathrm{s}}{V}
\right)y_{\mathrm{AA}}(r)\Delta V = \Delta  \overline{n}_\mathrm{s}(r)~,
\label{eq-yest} \end{eqnarray} 
when $n_\mathrm{t} \rightarrow \infty$. This is the same formula as if the
permissible insertions were actual particles though they are not.

\begin{figure}
\includegraphics[width=3.0in]{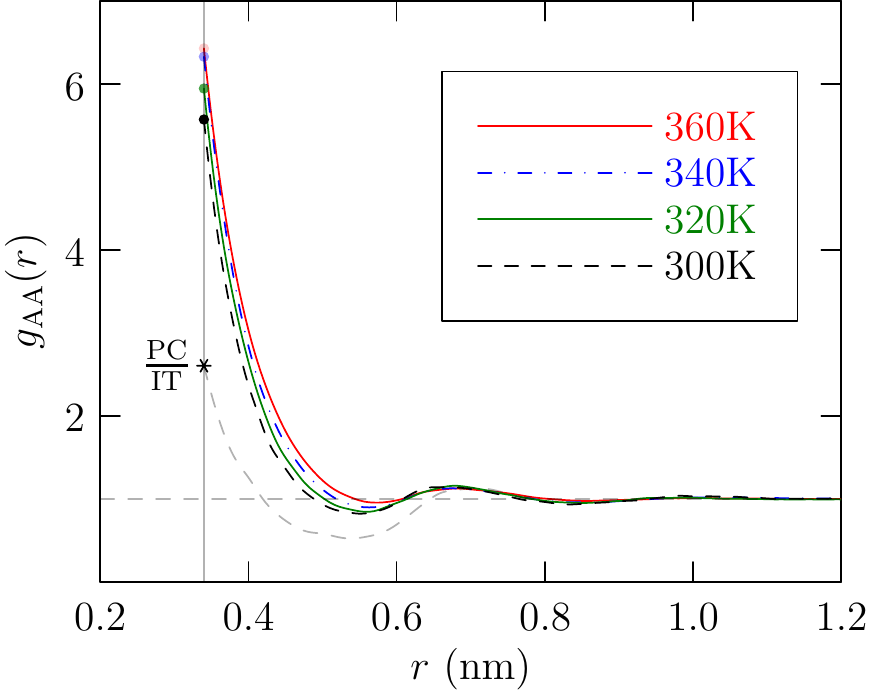}
\caption{Radial distribution functions for hard-sphere solutes in liquid
water at $p$ = 1~atm, and four different temperatures. The spheres have
van der Waals radius of 0.17~nm and distance-of-closest-approach to a
water oxygen atom of 0.31~nm. The prediction of the information theory (IT)
model \cite{ghumm96,Hummer98} at $T=300$~K is shown by the star
and the gray dashed curve. The contact value obtained matches the
Pratt-Chandler theory numerical result \cite{PRATTLR:Thehe}, and 
is labeled $\frac{\mathrm{PC}}{\mathrm{IT}}$.}
\label{fig:gHS} 
\end{figure}

The radial distribution function $g_{\mathrm{AA}}(r) = y_{\mathrm{AA}}(r)$ for
$r\ge 2~\times$ 0.17~nm = 0.34~nm, but is zero for $r < 0.34$~nm
(Fig.~\ref{fig:gHS}). The contact values, $g_{\mathrm{AA}}(r$ = 0.34~nm)
obtained are more than twice larger than the predictions of the Pratt-Chandler (PC) theory
\cite{PRATTLR:Thehe}. The contact values are higher for higher temperatures,
indicating stronger hydrophobic contact attractions at higher temperatures, 
agreeing with the results of Mancera, \emph{et al.,} and preceding
work discussed there \cite{Mancera:1997bv}. The contact values of the PC theory
also increase with $T$ but those increases are small \cite{PRATTLR:Thehe}, and
the PC contact values are sufficiently different from these numerical values
that the small increases are not interesting.

\section{Multi-Solute/Water Simulations to Assess Hydrophobic Interactions} Here
we discuss results from another assessment of hydrophobic interactions, namely,
 simulation of water with multiple hydrophobic solutes sufficiently aggressive
that they encounter each other enough to permit thermodynamic analysis of their
interactions \cite{Chaudhari:2015jq}.

\begin{figure}
\includegraphics[width=3.0in]{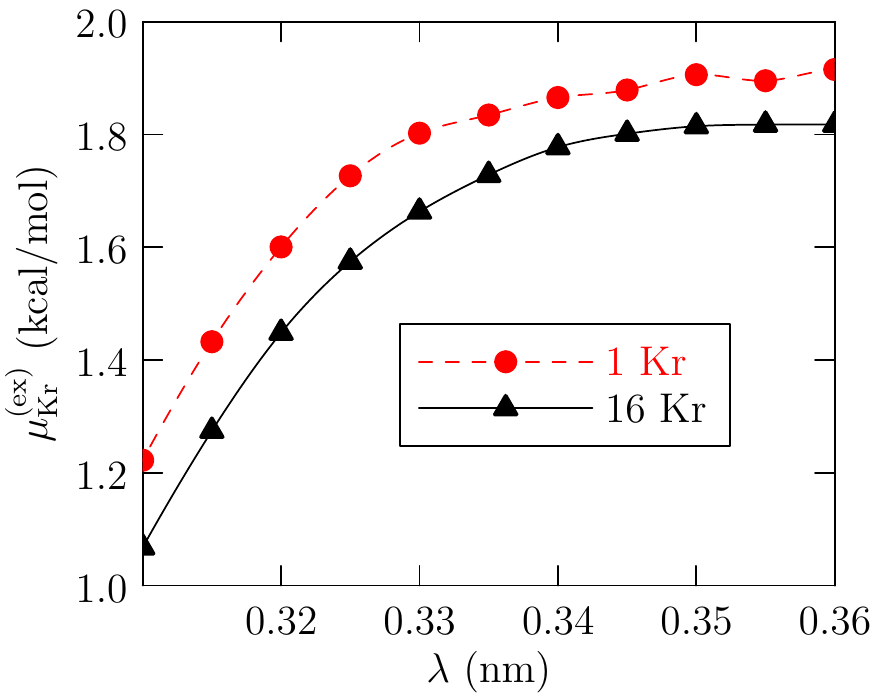}
\caption{Evaluations of hydration free energies
on the basis of quasi-chemical theory for a range of inner-shell boundaries
(0.31~nm   $< \lambda <$ 0.36~nm) for two Kr concentrations. 
$\mu_{\mathrm{Kr}}^{\mathrm{(ex)}}$ becomes insensitive 
in the range  0.34~nm $<\lambda<$ 0.36~nm.
The experimental value is 1.66 kcal/mol \cite{young}.}
\label{fig:six} 
\end{figure}

We will use molecular quasi-chemical theory (QCT) as our thermodynamic analysis tool
\cite{Rogers2012}. Results for two simulations, single Kr and multi-Kr, show
that the multi-Kr results for hydration free energy, $\mu_{\mathrm{Kr}}^{\mathrm{(ex)}}$, are distinctly
lower (FIG.~\ref{fig:six}). This already indicates that the hydrophobic
interactions are attractive, \emph{i.e.,} favorable. The significant difference
derives from slight reduction of the unfavorable packing contribution
identified by quasi-chemical theory. Two distinct further analyses then arrive
at concordant estimates of the osmotic second virial coefficient $B_2 \approx
-60~ \mathrm{cm}^3/\mathrm{mol}$ (attractive). Beyond the quasi-chemical theory
thermodynamic analysis, the observed Kr-Kr distributions were also  analyzed with
the extrapolation procedure of Kr\"{u}ger, \emph{et
al.}\cite{kruger2012kirkwood,schnell2013apply}. This approach provides a
convenient, theoretically neutral route to the evaluations of $B_2$ noted below.

Some thirty years ago, a focused molecular-dynamics study
\cite{WATANABEK:MOLSOT} estimated $B_2$ for Kr(aq) to be repulsive (positive).
Differences of the models treated and computational resources available probably
explain the difference of that previous evaluation with the present work.

\section{LMF/EXP Theory for Inclusion of Solute Dispersion Interactions 
for Ar Pair Hydrophobic Interactions}

\begin{figure}
\includegraphics[width=3.0in]{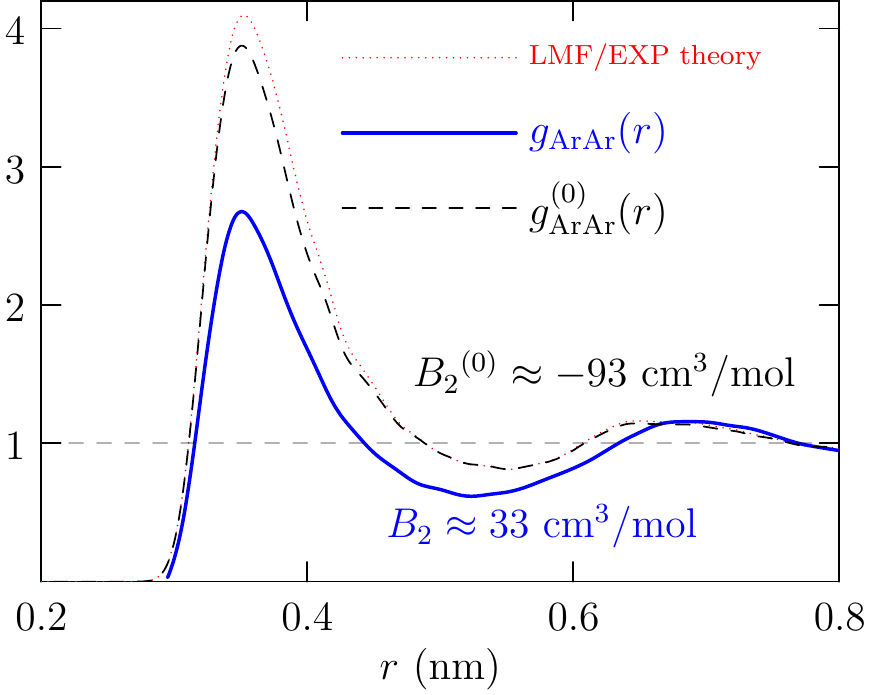}
\caption{For $T$ = 300~K, comparison of $g_{\mathrm{ArAr}}\left(r \right)$
(blue-solid) to the result for the reference system $g_{\mathrm{ArAr}}^{(0)}
\left(r \right)$ (LJ repulsions, black-dashed) and the LMF/EXP approximation
(red-dotted).  Solute attractive forces reduce the contact pair correlations and
thus weaken hydrophobic bonds. Note also the significantly different behavior of
$g_{\mathrm{ArAr}}^{(0)} \left(r \right)$ and $g_{\mathrm{ArAr}}\left(r \right)$
in the \emph{second shell}. Those differences suggest more basic structural
changes driven by attractive interactions.}
\label{fig:gAALMFCompare} 
\end{figure}

With the hard-sphere results of FIGs.~\ref{fg:gAO} and \ref{fig:gHS}, we
proceed further to discuss hydrophobic interactions involving further 
realistic interactions. Interactions $u_{\mathrm{AO}}$ and $u_{\mathrm{AA}}$ are
presented for analysis with A=Ar in the example above, and here
we consider solute interactions of Lennard-Jones type. As usual
\cite{mcquarrie12statistical}, these interactions are separated into a reference
part that describes all the repulsive forces, $u^{(0)}$, and a remainder
$u^{(1)}$.  We suppose that results corresponding to the reference system are
separately available, \emph{e.g.,} from direct numerical simulation such as
FIG.~\ref{fig:gAALMFCompare}. Then 
\begin{multline}
-\ln 
\left\lbrack 
\frac{g_{\mathrm{ArAr}}\left(\vec{r}\right)}{g_{\mathrm{ArAr}}^{(0)} \left(\vec{r} \right) }\right\rbrack  
\approx  
\beta u_{\mathrm{ArAr}}^{(1)}\left(\vec{r} \right)  \\
 +  \int 
h_{\mathrm{ArO}}\left(\vec{r}^\prime\right)
\rho_{\mathrm{O}}\beta u_{\mathrm{OAr}}^{(1)}\left(\vert\vec{r}^\prime - \vec{r}\vert\right)
\dif \vec{r}^\prime .
\label{eq:LMFapplication}
\end{multline}
is the simple theory to be tested. That theory is known as local 
molecular field theory (LMF).\cite{Rodgers:2008fd} 
The first term on the right of
Eq.~(3) builds-in the direct Ar-Ar attractive forces. The
second term supplies mean attractive forces from interaction of the solvent with
the solute. With $h_{\mathrm{ArO}} = g_{\mathrm{ArO}} -1$ conveniently taken
from routine simulation (FIG.~\ref{fg:gAO}), Eq.~(3) is
set so that the radial distribution functions on the left of
Eq.~(3) each approach one (1) at large separation. 

A concise rederivation \cite{Chaudhari:2016gm} of the approximate
Eq.~(3) emphasizes the basic concepts of the van~der~Waals
theories of liquids, and is persuasive on that basis. The corresponding
theory for $h_{\mathrm{ArO}}\left(\vec{r}\right)$ is
\begin{multline}
-\ln 
\left\lbrack 
\frac{g_{\mathrm{ArO}}\left(\vec{r}\right)}{g_{\mathrm{ArO}}^{(0)} \left(\vec{r} \right) }\right\rbrack  
\approx  
\beta u_{\mathrm{OAr}}^{(1)}\left(\vec{r} \right)  \\
 +  \int 
h_{\mathrm{OO}}\left(\vec{r}^\prime\right)
\rho_{\mathrm{O}}\beta u_{\mathrm{OAr}}^{(1)}\left(\vert\vec{r}^\prime - \vec{r}\vert\right)
\dif \vec{r}^\prime ,
\label{eq:EXPapplication}
\end{multline}
with $h_{\mathrm{OO}}\left(\vec{r}\right)$ the observed OO correlation
function for pure water. Acknowledging closure approximations specific to
traditional implementations, this is just the EXP approximation
\cite{hansen1976theory} applied to this correlation problem long ago
\cite{PrattLR:Effsaf,PhysRevLett.80.4193}.

The functions $u_{\mathrm{AO}}$ and $u_{\mathrm{AA}}$ are distinct, and the
interesting possibilities of Eq.~(3) lie in how
$u_{\mathrm{AO}}$ and $u_{\mathrm{AA}}$ contributions balance. In the present
application, these contributions balance closely in the relevant regime of
separations (FIG.~\ref{fig:gAALMFCompare}), predicting small changes due
to this balanced inclusion of attractive interactions.  Nevertheless, the observed
differences $g_{\mathrm{ArAr}}\left(r \right)$ and
$g_{\mathrm{ArAr}}^{\mathrm{(0)}}\left(r \right)$ are large in contact
geometries (FIG.~\ref{fig:gAALMFCompare}).

In this application, finally, we are forced again to the conclusions that
attractive interactions lead to substantial changes in the hydrophobic
interactions, and that the approximation Eq.~(3) does not
describe those changes well here. Work of long ago arrived at different
conclusions because accurate results for the reference system (FIGs.
\ref{fig:gHS} and \ref{fig:gAALMFCompare}) were not available for this analysis.

\begin{figure}
\includegraphics[width=3.0in]{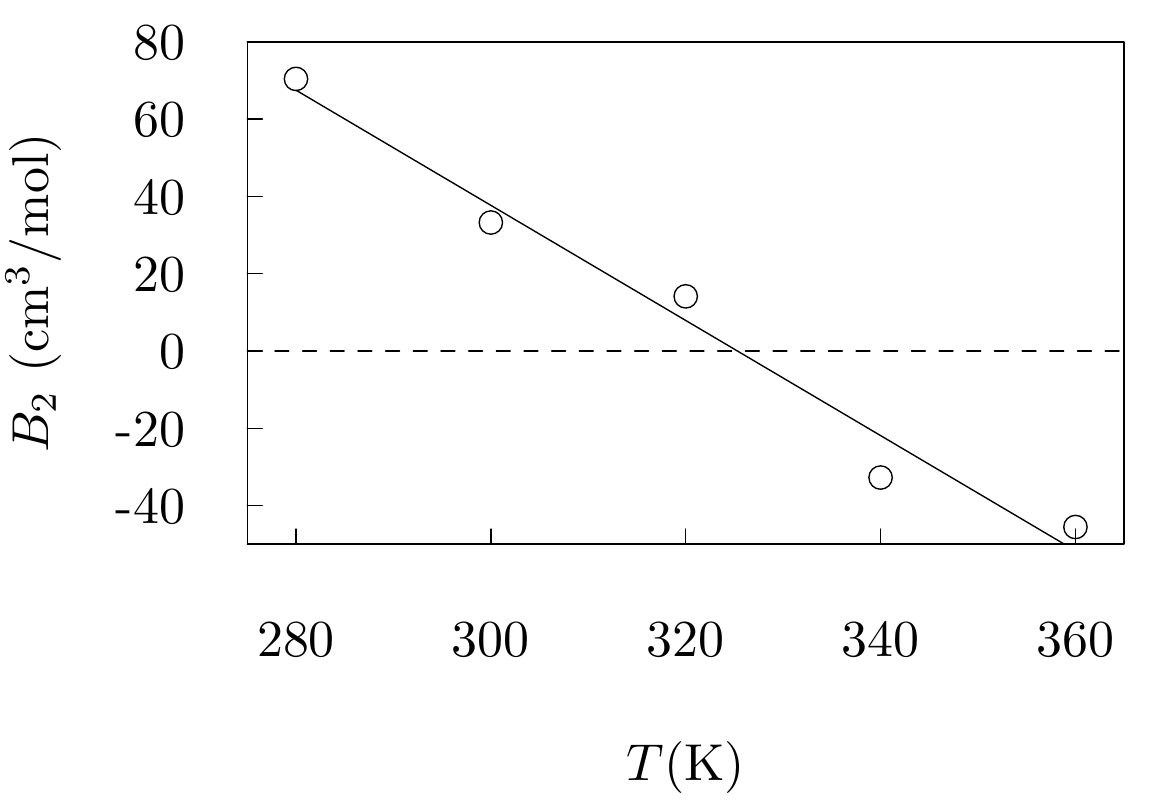}
\caption{Numerical values of the osmotic second virial
coefficients, $B_2$, obtained from the radial distribution functions 
using the extrapolation procedure of Kr\"{u}ger, \emph{et
al.}\cite{kruger2012kirkwood,schnell2013apply},
are nicely consistent with important recent results of Koga \cite{Koga:2013el}
and of Ashbaugh, \emph{et al.}\cite{Ashbaugh:2015cxa} }
\label{fig:B2TTikZ} 
\end{figure}

The second osmotic virial coefficient,
$B_2$, becomes more attractive with increasing temperature below $T$ = 360~K
(FIG.~\ref{fig:B2TTikZ}). This
behavior is consistent with important recent results of Koga \cite{Koga:2013el}
and of Ashbaugh, \emph{et al.}\cite{Ashbaugh:2015cxa} With attractive
interactions in play, $B_2$ can change from positive to negative values with
increasing temperatures. This is consistent also with historical work
\cite{WATANABEK:MOLSOT} that $B_2 \approx 0$ for intermediate cases.

\section{Discussion}

Changing purely repulsive atomic interactions to include realistic attractions
\emph{weakens} primitive hydrophobic bonds (FIG.~\ref{fig:gAALMFCompare}).
According to the LMF/EXP theory\cite{PrattLR:Effsaf,Asthagiri:2008in}, the
hydration environment competes with direct Ar-Ar attractive interactions
(Eq.~(3)). The outcome of that competition is sensitive to
the differing strengths of the attractive interactions. The earlier application
\cite{PrattLR:Effsaf} used the EXP approximation to analyze the available Monte
Carlo calculations on atomic LJ solutes in water \cite{PANGALIC:AMCs}. That
theoretical modeling found modest effects of attractive interactions, and
encouraging comparison with the Monte Carlo results. This application of the LMF
theory (Eq.~(3)) again predicts modest effects of attractive
interactions, but the net comparison from the simulation results shows big
differences. The alternative outcome is due to the fact that the earlier
application used the PC theory for the reference system
$g_{\mathrm{ArAr}}^{(0)}\left( r\right)$. We now know that approximation is not
accurate here \cite{chaudhari2013molecular}, despite being the only theory
available. Here the LMF theory (Eq.~(3)) predicts
modest-sized changes, also opposite in sign to the observed changes. Note
further that $g_{\mathrm{ArAr}}^{(0)}\left( r\right)$ and
$g_{\mathrm{ArAr}}\left( r\right)$ differ distinctively in the second hydration
shell, and those differences suggest more basic structural changes driven by
attractive interactions.

Earlier theoretical studies featured $\left\langle \varepsilon \vert r,
n_\lambda=0\right\rangle$, a central object in QCT for the present problem
\cite{Asthagiri:2008in}. A more accurate evaluation would involve $n$-body
($n>2$) correlations, even if treated by superposition approximations
\cite{Ashbaugh:2005jh}. Detailed treatment of the Ar$_2$ diatom geometry is the
most prominent difference between that QCT approach and the present LMF theory
(Eq.~(3)). Nevertheless, a full QCT analysis of these
differences is clearly warranted and should be the subject of subsequent study.

Finally, since attractions make large, masking contributions, tests of the PC
theory \cite{PANGALIC:AMCs,ghumm96} against results with realistic attractive
interactions should address the role of attractive interactions, which were not
included in the PC theory.

\section{Acknowledgement} Sandia National Laboratories (SNL) is a multiprogram laboratory operated by Sandia
Corporation, a Lockheed Martin Company, for the U.S. Department of Energy's
National Nuclear Security Administration under Contract No. DE-AC04-94AL8500.
The financial support of Sandia's LDRD program and the Defense Threat
Reduction Agency (DTRA) is gratefully acknowledged. The work was
performed, in part, at the Center for Integrated Nanotechnologies, an Office of Science User Facility operated for the U.S. DOE's Office of Science by Los Alamos National Laboratory (Contract DE-AC52-06NA25396) and SNL.

\bibliographystyle{achemso}


\providecommand{\latin}[1]{#1}
\providecommand*\mcitethebibliography{\thebibliography}
\csname @ifundefined\endcsname{endmcitethebibliography}
  {\let\endmcitethebibliography\endthebibliography}{}

\end{document}